\documentclass[aps,prd,nofootinbib,showpacs,twocolumn]{revtex4}
\usepackage{graphicx}

\begin{document}
\title{An entirely analytical cosmological model}
\author{Sandro Silva e Costa}
\affiliation{Centro de Ci\^{e}ncias Naturais e Humanas, Universidade Federal do ABC \\
Rua Santa Ad\'{e}lia, 166 --  Santo Andr\'{e} -- SP, 09210-170, Brazil}
\email{sandro.costa@ufabc.edu.br}
\date{\today}

\begin{abstract}
The purpose of the present study is to show that in a particular cosmological model, with an affine equation of state, one can obtain, besides the background given by the scale factor, Hubble and deceleration parameters, a representation in terms of scalar fields and, more important, explicit mathematical expressions for the density contrast and the power spectrum. Although the model so obtained is not realistic, it reproduces features observed in some previous numerical studies and, therefore, it may be useful in the testing of numerical codes and as a pedagogical tool.
\end{abstract}


\maketitle

Several studies about candidates for the role of a unified dark matter can be found in the literature \cite{linear}-\cite{alternative}. However, very few entirely analytical results are known, specially in the case of the evolution of perturbations. Here, the purpose is to show that, by choosing a certain equation of state, it is possible to have a completely analytical model, where one can obtain explicit mathematical expressions for the density contrast $\delta$ and, therefore, for the power spectrum $P\left(k\right)$, since they are related by the expression \cite{Peebles}
\begin{equation}
P\left(k\right)=\left|\delta\left(k\right)\right|^2\,.
\end{equation} 
Although not able to describe the current observations, the mathematical results shown here reproduce features observed in some numerical studies and, as such, may be useful in the testing of numerical codes and as a pedagogical tool.

The specific model studied in this text is characterized by an affine equation of state, where the relation between the pressure $p$ and the energy density $\rho$ of a cosmological fluid is\footnote{Throughout all the text natural units are used, i.e., $c=G=\hbar=k_B=1$.}
\begin{equation}
p=p_0+\alpha\rho\,,
\label{affine}
\end{equation}
where $p_0$ and $\alpha$ are constants. Such equation is a particular case of the quadratic model 
\begin{equation}
p=p_0+\alpha\rho+\beta\rho^2\,,
\end{equation}
with $\beta$ being a constant, presented in the literature as a possible simple phenomenological candidate for the role of a unified dark matter \cite{Ananda}. In the affine equation of state, eq. (\ref{affine}), the constant term in the pressure can be seen as due to a cosmological constant, to a constant bulk viscous pressure or to an effective pressure associated to the phenomenon of particle creation \cite{Martin}. This same kind of equation can be obtained from the equation of state defining the modified Chaplygin gas \cite{Benaoum}-\cite{mcg2},
\begin{equation}
\label{Chaplygin}
p=\left(\gamma-1\right)\rho-M\rho^{-\mu}\,,
\end{equation}
where $\gamma$, $M$ and $\mu$ are free parameters, by simply putting $\mu=0$, $M=-p_0$ and $\gamma=1+\alpha$. Since some basic results for the modified Chaplygin gas are available in the literature, it is useful to adopt such representation of the affine fluid.
  
Using the conservation of energy in an adiabatically expanding universe,
\begin{equation}
\frac{d\rho}{p+\rho}=-3\frac{da}{a}\,,
\end{equation}
where $a$ is the scale factor, one obtains, from eq. (\ref{Chaplygin}),
\begin{equation}
\rho=\left[A+\left(B-A\right)a^{-3\gamma\left(1+\mu\right)}\right]^{\frac{1}{1+\mu}}\,,
\end{equation}
where $A\equiv M/\gamma$ and $B\equiv\rho_0^{1+\mu}$, with $\rho_0$ being the present value of the energy density of the universe.
Substituting in this last result the value $\mu=0$ 
and using the Friedmann equation for a flat space,
\begin{equation}
\label{Friedmann}
\left(\frac{\dot{a}}{a}\right)^2=\frac{8\pi}{3}\rho\,,
\end{equation}
one obtains
\begin{equation}
\label{scalefactor}
a\left(t\right)=\left(\frac{B-A}{A}\right)^{\frac{1}{3\gamma}}\sinh^{\frac{2}{3\gamma}}\left[\frac{3\gamma}{2}\left(\frac{8\pi}{3}A\right)^{\frac12}\,t\right]\,,
\end{equation}
and, therefore, the Hubble parameter is
\begin{equation}
\label{Hubble}
H=\left(\frac{8\pi}{3}A\right)^{\frac12}\coth\left[\frac{3\gamma}{2}\left(\frac{8\pi}{3}A\right)^{\frac12}\,t\right]\,.
\end{equation}
The acceleration of the universe, given by the second Friedmann equation, is then
\begin{equation}
\frac{\ddot{a}}{a}=-\frac{4\pi}{3}\left(\rho+3p\right)
=-\frac{4\pi}{3}\left[\left(3\gamma-2\right)\rho-3M\right]\,.
\end{equation}
From these results is easy to see that the deceleration parameter is
\begin{equation}
q\equiv\frac{\ddot{a}}{aH^2}=\frac{3\gamma}{2}\frac{1}{\cosh^2\left[\frac{3\gamma}{2}\left(\frac{8\pi}{3}A\right)^{\frac12}\,t\right]}-1\,.
\end{equation}

Another interesting result that one can readily obtain is the representation of the model in terms of a time-dependent homogeneous scalar field \cite{scalar}. It is easy to show that putting
\begin{equation}
\rho=\frac{\dot{\varphi}^2}{2}+V\left(\varphi\right)\,,\,\,
p=\frac{\dot{\varphi}^2}{2}-V\left(\varphi\right)\,,
\end{equation}
where $\varphi\left(t\right)$ is a scalar field with an associated potential $V\left(\varphi\right)$, one has
\begin{equation}
\varphi=\varphi_0+\frac{1}{\sqrt{6\pi\gamma}}\ln\left\{\coth\left[\frac{3\gamma}{4}\left(\frac{8\pi}{3}A\right)^\frac12 t\right]\right\}
\end{equation}
and
\begin{eqnarray}
V\left(\varphi\right)&=&\frac{A}{2}\cosh^2\left[\sqrt{6\pi\gamma}\left(\varphi-\varphi_0\right)\right]
\nonumber\\
&&\times\left\{2-\gamma\tanh^2\left[\sqrt{6\pi\gamma}\left(\varphi-\varphi_0\right)\right]\right\}\,\,,
\end{eqnarray}
where $\varphi_0$ is a constant.

Beyond the background given by the scale factor, Hubble and deceleration parameters, one important way of characterizing a cosmological model consists of studying the evolution of perturbations on it. Without considering the ``averaging problem'' \cite{linear}, one can write the equation for the perturbations (cf. equation 4.122 from Padmanabhan \cite{Padmanabhan}),
\begin{eqnarray}
\frac{d^2\delta}{da^2}+\frac{3-15\omega+6v^2}{2a}\frac{d\delta}{da}+\frac{k^2v^2\delta}{H^2a^4}
\nonumber\\=\frac{3\delta}{2a^2}\left(1-6v^2-3\omega^2+8\omega\right)\,,
\end{eqnarray}
where $\delta$ is the density contrast, 
$\omega\equiv p/\rho$, 
$v^2\equiv\frac{\partial p}{\partial\rho}$ 
and $k$ is the wavenumber of the Fourier mode in consideration. Therefore, in the family of models considered here, 
\begin{equation}
\label{omega}
\omega=\gamma-1-M\rho^{-1}
=v^2-M\rho^{-1}\,,
\end{equation}
i.e., $v^2$ is a constant, which may be negative if $\gamma<1$.

Using in the equation for the perturbations the variable
\begin{equation}
x=-\left(\frac{B-A}{A}\right)a^{-3\gamma}=-\sinh^{-2}\left[\frac{3\gamma}{2}\left(\frac{8\pi}{3}A\right)^{\frac12}\,t\right]\,,
\end{equation}
and the hypothesis
\begin{equation}
\delta=x^\frac{1}{\gamma}\left(1-x\right)^{-\frac12}F\left(x\right)\,,
\end{equation}
one can obtain the equation
\begin{eqnarray}
\label{eqF}
\left(1-x\right)x\frac{d^2F}{dx^2}+\left[\frac{1}{3\gamma}-\left(\frac{3}{2}+\frac{1}{3\gamma}\right)x\right]\frac{dF}{dx}\nonumber\\=-\frac{\left(\gamma-1\right)k^2}{24\pi\gamma^2A}\left(-\frac{A}{B-A}\right)^{\frac{2}{3\gamma}}x^{\frac{2}{3\gamma}-1}F\,.
\end{eqnarray}

A solution of the equation (\ref{eqF}) is easily found only for a few specific cases: $k=0$, $\gamma=1$ and $\gamma=2/3$. The last one is more interesting because then one obtains a density contrast which depends on the wavenumber $k$. Explicitly, the solution for this case is 
\begin{equation}
\delta=c_1\delta_1+c_2\delta_2\,,
\label{delta0}
\end{equation}
with $c_1$ and $c_2$ being arbitrary constants, 
\begin{equation}
\delta_1=\frac{x^{\frac{3}{2}}}{\left(1-x\right)^\frac12}\,_2F_1\left[\frac12+\sqrt{\tilde{k}^2+\frac14},\frac12-\sqrt{\tilde{k}^2+\frac14};\frac12;x\right]
\label{delta1}
\end{equation}
and
\begin{equation}
\delta_2=\frac{x^{2}}{\left(1-x\right)^\frac12}\,_2F_1\left[1+\sqrt{\tilde{k}^2+\frac14},1-\sqrt{\tilde{k}^2+\frac14};\frac{3}{2};x\right]\,,
\label{delta2}
\end{equation}
where $\tilde{k}^2\equiv k^2/\left[32\pi\left(B-A\right)\right]$. Another equivalent way of writing such result is 
\begin{equation}
\delta=\frac{x^{\frac32}}{1-x}\left[c_1\cos\left(\nu\arcsin x^{\frac12}\right)+c_2\sin\left(\nu\arcsin x^{\frac12}\right)\right]\,,
\label{delta3}
\end{equation}
where $\nu\equiv\left(1+4\tilde{k}^2\right)^{1/2}$. 



There is yet a third way of writing the solution given in equations (\ref{delta0}), (\ref{delta1}) and (\ref{delta2}), which is obtained from the aplication on equation (\ref{delta3}) of the definitions of the inverse trigonometric function arcsine \cite{Gradshteyn} and of the variable $x$. Then one has
\begin{equation}
\delta=c_1^\prime\delta_++c_2^\prime\delta_-\,,
\label{delta4}
\end{equation}
where $c_1^\prime$ and $c_2^\prime$ are arbitrary constants,
\begin{equation}
\delta_\pm=\frac{1}{\sinh\bar{t}\cosh^2\bar{t}}\left(\frac{\cosh\bar{t}\mp 1}{\sinh\bar{t}}\right)^{\nu}\,,
\label{delta4a}
\end{equation}
and where $\nu\equiv\left(1+4\tilde{k}^2\right)^{1/2}$ and $\bar{t}=\left(4\pi M\right)^{1/2}t$.
This result can be obtained directly from the equation for the density contrast using $t$ as variable \cite{Padmanabhan},
\begin{eqnarray}
\label{Pad1}
\ddot{\delta}+H\left[2-3\left(2\omega-v^2\right)\right]\dot{\delta}
+k^2\frac{v^2}{a^2}\delta\nonumber\\=\frac{3}{2}H^2\left(1-6v^2-3\omega^2+8\omega\right)\delta\,.
\end{eqnarray}
Using eqs. (\ref{Friedmann}), (\ref{scalefactor}), (\ref{Hubble}) and (\ref{omega}), with $\gamma=2/3$, changing the variable from $t$ to $\bar{t}$, and choosing
\begin{equation}
\delta=\frac{G\left(\bar{t}\right)}{\sinh\bar{t}\cosh^2\bar{t}}\,,
\end{equation}
one obtains the equation
\begin{equation}
\frac{d^2G}{d\bar{t}^2}+\coth\bar{t}\frac{dG}{d\bar{t}}-\frac{\nu^2}{\sinh^2\bar{t}}G=0\,,
\end{equation}
which admits as solutions the toroidal functions $P_{0}^{\nu}\left(\cosh\bar{t}\right)$ and $Q_{0}^{\nu}\left(\sinh\bar{t}\right)$ \cite{Gradshteyn} or, simply, the result given by equations (\ref{delta4}) and (\ref{delta4a}).

\begin{figure}
\centerline{\includegraphics[scale=0.9]{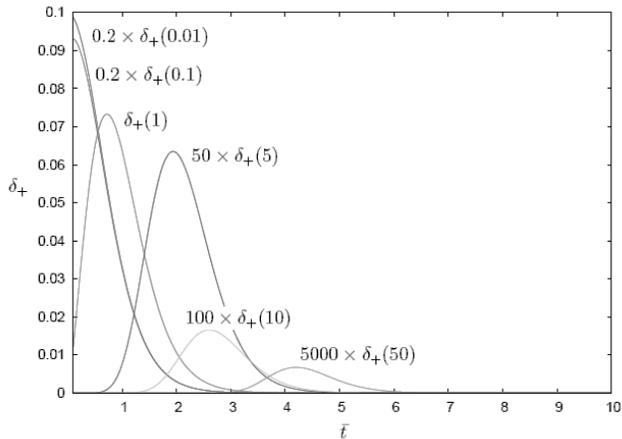}}
\caption{Curves showing the variation in time of the function $\delta_+\left(\tilde{k}\right)$ for some values of $\tilde{k}$. All curves present peaks whose location and magnitude varies with $\tilde{k}$. Notice that the time variable is $\bar{t}=\left(4\pi M\right)^\frac12 t$, where $t$ is the usual cosmological time, with $M$ being a parameter of the model (see eq. (\ref{Chaplygin})).}
\label{deltaplus}
\end{figure}

\begin{figure}
\centerline{\includegraphics[scale=0.9]{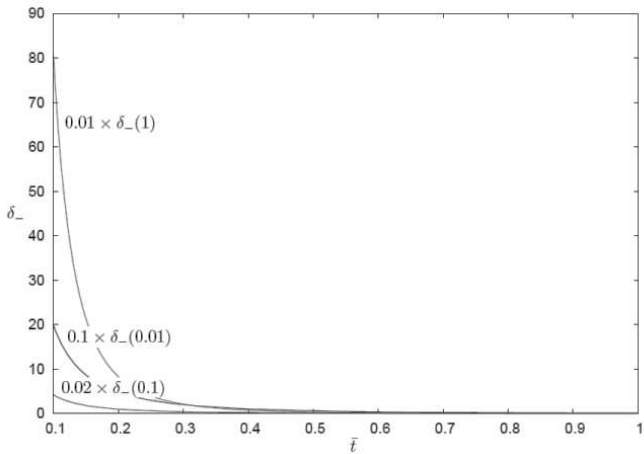}}
\caption{Curves showing the variation in time of the function $\delta_-\left(\tilde{k}\right)$ for some values of $\tilde{k}$. Notice that the time variable is $\bar{t}=\left(4\pi M\right)^\frac12 t$, where $t$ is the usual cosmological time, with $M$ being a parameter of the model (see eq. (\ref{Chaplygin})).}
\label{deltaminus}
\end{figure}

Figures \ref{deltaplus} and \ref{deltaminus} show the behaviour in time of the functions $\delta_+$ and $\delta_-$ for some values of $\tilde{k}$. It is easy to notice that $\delta_-$ represents rapidly decreasing modes, which dominate for smaller times, while $\delta_+$ represents modes with peaks which become the dominating ones for greater times. The presence of peaks in the density contrast is a feature shown in previous numerical studies \cite{GRG,Lazkoz}  and, therefore, confirmed here analitically.

\begin{figure}
\centerline{\includegraphics[scale=0.31]{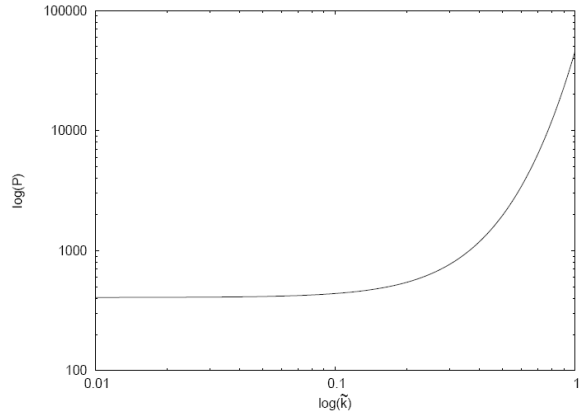}}
\caption{Graph showing the behaviour of the power spectrum $P$, as given by equation (\ref{eqP1}), for some values of $\tilde{k}$, at the time $\bar{t}=0.3$, chosen arbitrarily. Notice that the scale in the vertical axis is arbitrary.}
\label{P1}
\end{figure}

\begin{figure}
\centerline{\includegraphics[scale=0.31]{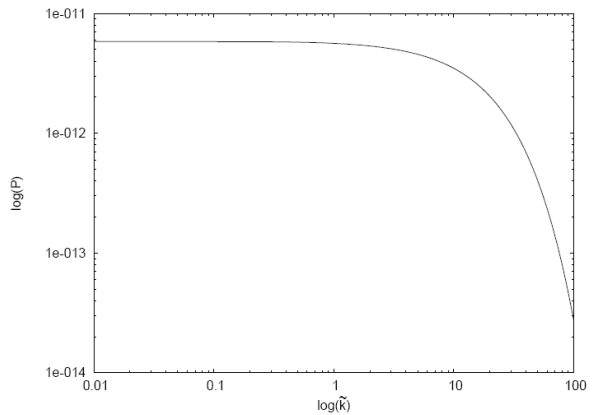}}
\caption{Graph showing the behaviour of the power spectrum $P$, as given by equation (\ref{eqP2}), for some values of $\tilde{k}$, at the time $\bar{t}=5$, chosen arbitrarily. Notice that the scale in the vertical axis is arbitrary.}
\label{P2}
\end{figure}

It is easy now to construct the power spectrum $P\left(k,t\right)$ associated to this solution, with two types of modes. Essentially, if the decaying mode is dominating one has
\begin{equation}
P\left(k,t\right)\approx c_1^2\left[\delta_-\left(t\right)\right]^2\,,
\label{eqP1}
\end{equation}   
while if the growing mode is dominant one has
\begin{equation}
P\left(k,t\right)\approx c_2^2\left[\delta_+\left(t\right)\right]^2\,.
\label{eqP2}
\end{equation}
Figures \ref{P1} and \ref{P2} show the power spectrum thus obtained for arbitrary times. 

It must be clear that none of the results shown here are consistent with the current observations. For example, they are not compatible, at the same time, with an initial scale invariant power spectrum ($P\propto k$) and the effectively measured power spectrum \cite{Cole}: if one sets the constants $c_1$ and $c_2$ (or $c^\prime_1$ and $c^\prime_2$) to give the correct initial conditions, one misses the target of real data, and vice-versa. Therefore, the model shown here is simply a toy model, or a mathematical possibility, but an interesting one because it has the peculiarity of yielding analytical results along all the way. Anyway, this may be seen as a practical example of the idea behind the famous Einstein's phrase ``As far as the laws of mathematics refer to reality, they are not certain, and as far as they are certain, they do not refer to reality'' \cite{Einstein}.

\section*{Acknowledgements} 
The author thanks M. Ujevic for fruitful discussions and a revision of the text.

\end{document}